\newcommand*{\krlastchange}{2022-10-12}

\documentclass[14pt,british]{extarticle}

\usepackage{babel}
\usepackage[utf8]{inputenc}

\usepackage{datetime2}

\DTMsetup{useregional}

\usepackage{fancyhdr}
\usepackage{url}

\usepackage{amsmath}
\usepackage{amsfonts}
\usepackage{framed}
\usepackage{graphicx}
\usepackage{caption}
\usepackage{subcaption}

\usepackage{titlesec}
\titleformat{\section}{\normalsize\bfseries}
   {\thesection}{1em}{}
\titleformat{\subsection}{\normalsize\bfseries}
   {\thesubsection}{1em}{}

\begin{document}


\DTMsavedate{krrevdate}{\krlastchange}


\newcommand*{\revdatefoot}
{\DTMtwodigits{\DTMfetchday{krrevdate}}-\DTMenglishshortmonthname{\DTMfetchmonth{krrevdate}}-\DTMfetchyear{krrevdate}}

\newcommand*{\revdatelong}{\DTMenglishmonthname{\DTMfetchmonth{krrevdate}} \DTMfetchday{krrevdate}, \DTMfetchyear{krrevdate}}

\newcommand*{\revdateyyyymmdd}
{\DTMfetchyear{krrevdate}\DTMtwodigits{\DTMfetchmonth{krrevdate}}\DTMtwodigits{\DTMfetchday{krrevdate}}}


\setlength{\parindent}{0in}
\setlength{\parskip}{0.4in plus0.2in minus0.2in}
\setlength{\voffset}{0in}
\setlength{\topmargin}{0in}
\setlength{\headheight}{0in}
\setlength{\headsep}{0in}
\setlength{\footskip}{0.5in}

\pagestyle{plain}

\noindent

\inputencoding{utf8}

\bibliographystyle{plain}


\long\def\symbolfootnote[#1]#2{\begingroup%
\def\thefootnote{\fnsymbol{footnote}}
     \footnote[#1]{#2}\endgroup}

\newcommand*{\Schrodinger}{Schrödinger\hspace{0em}}










\begin{center}
{\large\textbf{Lambert W Lines and \\
  Finite Square Well Sensors}}
 ====================================\\
 \, \\
Ken Roberts,
Najeh Jisrawi, 
J. Jeyasitharam, \\
Shreyas Suresh, 
P. C. Deshmukh 
and S. R. Valluri \\
\revdatelong\footnote{KR, NJ and SRV are
at Physics and Astronomy, Western University, London, Canada. \\
JJ, SS, and PCD are at IITTp, Indian Institute of Technology at Tirupati, India. \\
Emails for correspondence:
krobe8@uwo.ca, valluri@uwo.ca}
\end{center}



\pagestyle{fancy}
\fancyhf{}
\renewcommand{\headrulewidth}{0pt}
\fancyfoot[L]{Page \thepage}
\fancyfoot[R]{\revdatefoot}

\textbf{Abstract}

The bound state energies of a 1-dimensional finite
quantum square well (FSW) can be determined using a
geometric method, involving a smooth mapping between 
two copies of the complex plane.   
The method allows one to identify particular strengths
of the FSW at which the system can become unusually
sensitive to changes in the well depth or geometry.
In the present paper we explore that sensitivity,
and exhibit a 3-D visualization of the solutions.




\section{Introduction}
\label{sectintro}

We previously described a geometric method for 
determining the bound state energies of a 
1-dimensional finite quantum square 
well\cite{RV2014,RV2017}.
The method involves a smooth mapping between
two copies of the complex plane, and provides one
with two alternative visualizations of the solutions.
In section \ref{sect2d} of this paper we will summarize
that method, in order to establish our terminology.

There are certain strengths of an FSW which are
of particular interest, because the number of bound
states of the FSW is quite sensitive to the parameters
of the well at those strengths.  A slight change of the
potential depth of the well, or of the well geometry
due to flexure, can cause meaningful change in the
bound states. 
In section \ref{sect3d}, we extend the 2-D
representations of the FSW to a third dimension,
in order to visualize the changes in the solution
sets as the strength of the FSW is altered.
That can provide one with guidelines for the design
of sensors.

In future work, we plan to consider
some FSW-like systems whose mathematical
representations involve equations similar to 
those which appear in FSW models.



\section{Geometric Solution of FSW Systems}
\label{sect2d}

In this section, we will introduce the concepts of 
an FSW-like system, 
Lambert lines, and
the two-planes method.

\subsection{What is an FSW-Like System?}
\label{sectfswsystem}

Many quantum mechanics textbooks describe the mathematics
of the 1-dimensional finite square well, and present equations
whose solution will determine the energy levels of the bound
states of the FSW.\footnote{
  See, for example, 
  D. Bohm\cite{DBohm1951}, section 11.10;
  Bransden and Jochain\cite{BJ1989}, section 4.6;
  Davies and Betts\cite{DB1994}, section 2.4; or 
  Griffiths\cite{Griffiths2005}, section 2.6.
}
After simplification and the introduction of dimensionless
coordinates, the problem comes down to finding the
simultaneous solutions of two equations.
The equations can be written in real variables $u,v$,
and a dimensionless positive parameter $R$.
One equation is 
\begin{eqnarray}
  \label{equvr}
    u^2 + v^2 &=& R^2 
\end{eqnarray}
The solutions are thus constrained to lie on the
circumference of a circle of radius $R$ in the $(u,v)$-plane.
The other equation is one of the two following:
\begin{eqnarray}
  \label{equvtan}
  u &=& v \, \tan(v) \,\,\,\,\,\,\, \textrm{(for even bound states)} \\
  \label{equvcot}
  u &=& -v \, \cot(v) \,\,\, \textrm{(for odd bound states)}
 \end{eqnarray}
 
 We expect, in future papers, 
 to consider models of other physical systems 
 which lead to similar simultaneous equations.
 For convenience, we will refer to all such systems 
 as  {\textbf{FSW-like systems}}.
 The system may be classical rather than quantum.  
 The distinguishing property of an 
 FSW-like system is that it involves, 
 when suitable coordinates are introduced, 
 the simultaneous solution of one or more
 equations which have a similar structure to
 equations (\ref{equvtan}) or (\ref{equvcot})

\subsection{What are Lambert Lines?}

Instead of considering the solutions of
one of equations (\ref{equvtan}) or (\ref{equvcot})
in the $(u,v)$-plane, let's consider
the solutions in a complex $w$-plane,
where $w = u + \mathbf{i}v$.
Moreover, define another complex
plane, the $z$-plane, related to the
$w$-plane by the map $z = F(w)$, 
given by
\begin{eqnarray}
    z = F(w) &=& w \, \exp(w).
\end{eqnarray}
Introduce real and imaginary coordinates
in the $z$-plane, 
so that $z = x + \mathbf{i} y$.

In terms of the $(u,v)$ and $(x,y)$ coordinates,
we have
\begin{eqnarray}
  z &=& x + \mathbf{i} y \\
    &=& F(w) \\
    &=& F(u + \mathbf{i}v) \\
    &=&
       \big(u + \mathbf{i}v\big) \, 
       \exp(u + \mathbf{i}v) \\
    &=& \big(u + \mathbf{i}v\big) \, 
       \exp(u) \, \exp(\mathbf{i}v) \\  
    &=& \big(u + \mathbf{i}v\big) \, 
       \exp(u) \, 
       \big[ \cos(v) + \mathbf{i}\sin(v) \big] \\
    &=& \exp(u) \, 
       \big[ u \, \cos(v) - v \sin(v) \big] \\
       \nonumber
       &\,& + \mathbf{i} \, \exp(u) \, 
       \big[ v \, \cos(v) + u \sin(v) \big] \\
    \label{eqzuv}
    &=& \exp(u) \, \cos(v) 
       \big[ u - v \tan(v) \big] \\
       \nonumber
       &\,& + \mathbf{i} \, \exp(u) \, \sin(v)
       \big[ v \, \cot(v) + u \big]  
\end{eqnarray}

Hence:  The solutions to $u = v \tan(v)$
are exactly the points of the $w$-plane
for which $z = F(w)$ is pure imaginary.
The solutions to $u = -v \cot(v)$
are exactly the points of the $w$-plane 
for which $z = F(w)$ is pure real.

The inverse of the $z = F(w)$ map
is known as the Lambert $W$
function\footnote{
  For background about the Lambert $W$
  function, see 
  \cite{Corless1996, Valluri2000},
  section 4.13 of \cite{Olver2010},
  and \cite{Mezo2022}.
  }
Lambert $W$ is a 
multi-branch complex function,
usually denoted by a capital letter $W$.
To say that $w = W(z)$,
for some branch of the $W$ function,
means precisely that $w \exp(w) = z$.

The {\textbf{Lambert Lines}} are the
curves in the $w$-plane for which 
$z = w \exp(w)$ is either pure imaginary
or pure real.
The imaginary Lambert lines are 
the solutions of $u = v \tan(v)$, 
and the real Lambert lines are 
the solutions of $u = -v \cot(v)$.

\begin{figure}[ht]
	\centering
	\includegraphics[width=1.0\textwidth]{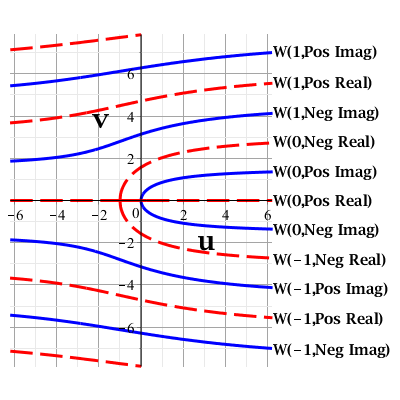}
     \caption{Lambert Lines in the $w$-Plane.
     The imaginary Lambert lines are
     solid curves drawn in blue.
     The real Lambert lines are
     dashed curves drawn in red.
     The real $u$-axis ($v = 0$) 
     of the $w$-plane has a dashed red curve
     overlaid upon it, corresponding
     to the real values of the Lambert $W$ function
     of a real argument.  Those values arise from
     branches 0 and -1 of Lambert $W$, for a
     real argument $z = x \ge -1/e$. \\
     ------------------------------------------------------------------------------------------} 
	\label{fig01}
\end{figure}

Figure \ref{fig01} shows the Lambert lines.
The imaginary Lambert lines are solid
curves shown in blue, 
and the real Lambert lines are dashed
curves shown in red.
The various Lambert lines have been labelled
to show their branch numbers and which axial
ray of the $z$-plane that Lambert line 
corresponds to.  For instance, the solid
blue Lambert line near the bottom of figure
\ref{fig01}, 
which is labelled $W(-1,\text{Neg Imag})$,
is obtained from the negative imaginary
axial ray of the $z$-plane (that is, the 
points $z = x + \mathbf{i}y$ with $x = 0$
and $y < 0$), by determining the curve
$w = W_{-1}(z)$ in the $w$-plane.
Here the $W_{-1}()$ notation indicates
that one is evaluating branch -1 of the
Lambert $W$ function.

\begin{figure}[ht]
	\centering
	\includegraphics[width=1.0\textwidth]{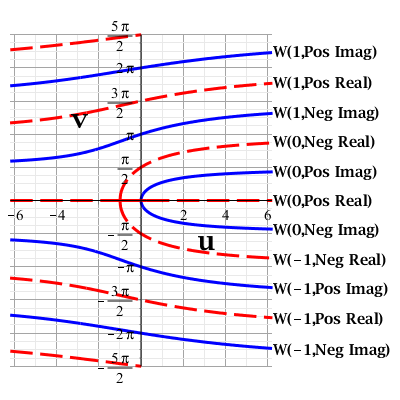}
     \caption{Lambert Line Intersections of Vertical $v$-Axis.
     This is the same as figure \ref{fig01},
     but vertical axis ($v$-axis) tick marks
     have been made at integer
     multiples of $\pi/2$.
     Imaginary Lambert lines (blue solid curves)
     intersect the $v$-axis
     at even multiples of $\pi/2$,
     and real Lambert lines (red dashed curves)
     intersect the $v$-axis
     at odd multiples of $\pi/2$. \\
     ------------------------------------------------------------------------------------------} 
	\label{fig02}
\end{figure}

Each Lambert line intersects the vertical
$v$-axis ($u = 0$) at an integer multiple of $\pi/2$.
Imaginary Lambert lines (blue) intersect the
$v$-axis at an even multiple of $\pi/2$,
and real Lambert lines (red) intersect the
$v$-axis at an odd multiple of $\pi/2$.
This is easier to see in figure \ref{fig02},
which is the same as figure \ref{fig01}, 
but with the vertical axis tickmarks at
integer multiples of $\pi/2$.

\begin{figure}[ht]
	\centering
      \includegraphics[width=1.0\textwidth]{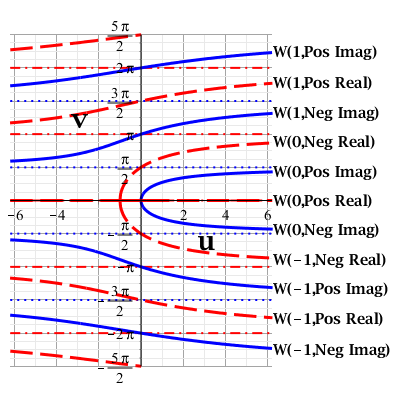}
	\caption{Lambert Line Asymptotes.
     The same as figure \ref{fig02},
     with the asymptotes of the Lambert
     lines included.
     The imaginary Lambert lines are 
     solid curves drawn in blue, 
     and their asymptotes are 
     horizontal dotted lines drawn in blue.
     The real Lambert lines are 
     dashed curves drawn in red,
     and their asymptotes are
     horizontal dot-dash lines drawn in red. \\
     ------------------------------------------------------------------------------------------} 
	\label{fig03}
\end{figure}

As the magnitude of $u$ goes to infinity,
the imaginary Lambert lines are asymptotic
to $v$ equal to an odd integer multiple of $\pi/2$, 
and the real Lambert lines are asymptotic
to $v$ equal to an even integer multiple of $\pi/2$.
In figure \ref{fig03} we have redrawn
figure \ref{fig02} to include the 
asymptotes of the Lambert lines.
The asymptotes of the imaginary Lambert lines
are drawn as blue dotted lines, and the 
asymptotes of the real Lambert lines are
drawn as red dash-dot lines.

\subsubsection{Comparison with Real Lambert $W$}

\begin{figure}[ht]
	\centering
      \includegraphics[width=1.0\textwidth]{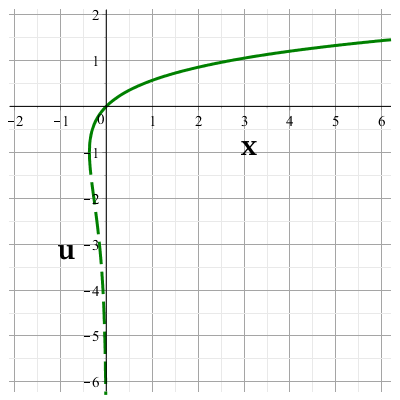}
	\caption{Real $u$ = Lambert $W$
	of Real Arg $x$.
	If $z = x$ has zero imaginary part,
	and if $w = u = W(x)$ is also real,
	then one can graph $u$ as a
	function of $x$, as in this figure.
	There are two branches of Lambert $W$
	which can have a real value of a
	real argument.
	Branch 0, the principal branch,
	has $u$ real provided $-1/e \le x$.
	That branch will have $u \ge -1$.
	It has been drawn here as a solid curve,
	in green.
	Branch -1 has $u$ real provided
	$-1/e \le x < 0$.
	That branch will have $u \le -1$.
	It has been drawn here as a dashed
	curve, in green.
	Note that this graph is in the $(x,u)$
	plane, and is a relationship between
	two real variables $x$ and $u$. \\
      ------------------------------------------------------------------------------------------} 
	\label{fig04}
\end{figure}

You may have seen previous graphs of the
Lambert $W$ function which look something
like figure \ref{fig04}.
That can cause some confusion, 
as that figure \ref{fig04} 
has a much different appearance 
from figure \ref{fig01}.

In each of figures \ref{fig01},
\ref{fig02} and \ref{fig03}
there are some real Lambert lines which are hard
to see, because they coincide
with the horizontal axis ($u$-axis)
in the $w$-plane.
Those are the inverse images of the real
axis ($x$-axis) of the $z$-plane,
for which the $w$ values are also real
(so that $v = 0$).
That is, in the relationships $z = w \, \exp(w)$
and its inverse $w = W(z)$, 
these particular Lambert lines restrict both
$z$ and $w$ to be real-valued.
In this situation, $w = u$ and $z = x$.
There are many scientific 
problems (for instance, solar cell models)
whose standard solution requires only 
the real-valued Lambert $W$ function 
$u$ of a real variable $x$.

Section 4.13 of the NIST Handbook\cite{Olver2010}
restricts its discussion to Lambert $W$ as a real-valued
function of a real variable.
Figure \ref{fig04} shows a graph
 of $u = W(x)$,
of course drawn in the $(x,u)$ plane.
There are two branches of Lambert $W$.
For $-1/e \le x$, if we take $u \ge -1$
then we get the branch called $W_p$
in \cite{Olver2010}.
And for $-1/e \le x < 0$, if we take $u \le -1$,
then we get the branch called $W_m$
in \cite{Olver2010}.
In the more standard terminology,
such as used in 
\cite{Corless1996, Valluri2000, Mezo2022}
for complex valued $w = W(z)$
Lambert $W$ of a complex argument $z$, 
the branch $W_p$ of \cite{Olver2010} is 
the real part of branch 0 
of the complex Lambert $W$ function,
and the branch $W_m$ of \cite{Olver2010}
is the real part of branch -1 
of the complex Lambert $W$ function.

How do the curves in the $(x,u)$ plane
of figure \ref{fig04}
appear in the $w$-plane and $z$-plane?
Since $y = 0$, we are mapping the
real axis of the $z$-plane to the $w$-plane.
And since $v = 0$, the value of the Lambert
$W$ function in the $w$-plane must also
be on the real axis of the $w$-plane.
Thus these real-valued Lambert line curves
correspond to branch 0 of the
Lambert $W$ function of a real variable
$z = x \ge -1/e$ and to branch -1 of
the Lambert $W$ function of a 
negative real variable $z = x$
which satisfies $0 > x \ge -1/e$.
In each case $w = u$ is real valued.
For branch 0, when $x \ge -1/e$,
we have $u \ge -1$ and $v = 0$.
For branch -1, when $-1/e \le x < 0$,
we have $u \le -1$ and $v = 0$.
Thus each of those partial Lambert lines
is drawn on top of the horizontal
axis of the $w$-plane in figures
\ref{fig01}, \ref{fig02} and \ref{fig03}.

\begin{figure}[ht]
	\centering
	\includegraphics[width=1.0\textwidth]{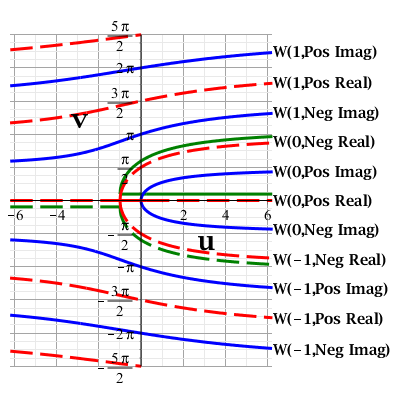}
     \caption{Branch 0 and -1 Real Lambert Lines.
     The real Lambert lines for branch 0
     (for all real values of $z = x$) have been
     overlaid on the standard $w$-plane
     Lambert lines diagram, 
     with the real Lambert lines
     for branch 0 drawn as a solid green curve,
     shifted up by $0.3$ to enable the green
     curve to be distinguished from the
     $v = 0$ abscissa of the $w$-plane.
     The real Lambert line for branch -1
     (for real values of $z = x < 0$) have been
     overlaid on the standard $w$-plane
     Lambert lines diagram,
     with the real Lambert line
     for branch -1 drawn as a dashed green curve,
     shifted down by $0.3$ to enable the green
     curve to be distinguished from the 
     abscissa of the $w$-plane. \\
     ------------------------------------------------------------------------------------------} 
	\label{fig05}
\end{figure}

To make the structure clearer, in 
figure \ref{fig05} we have
taken figure \ref{fig02} and
drawn, in the complex $w$-plane,
additional copies of the real Lambert lines
for the branches 0 and -1 of the
complex Lambert function.
In order that the real-valued portions
of those curves will be visible,
we have offset these additional copies
(drawn in green) from the standard
real Lambert lines (drawn in red)
for those two branches, branch 0 and
branch -1.

\textbf{Branch 0:}
We have used a solid green line
for branch 0, and added $0.3$ to the
value of $v$, so that we are graphing
\begin{eqnarray}
  w &=& u + \mathbf{i}v 
                + \mathbf{i}[\text{offset}] \\
     &=& W_{0}(z) + 0.3\mathbf{i}
\end{eqnarray}
for real $z = x$.
This displays the $w$-plane images
of both of the real 
axial rays of the $z$-plane.
When $-1/e \le x$, 
the branch 0 real Lambert line
(solid green line) is horizontal,
shifted 0.3 above the abscissa,
and that portion of the Lambert
line has $u \ge -1$.
When $x < -1/e$, the $w$ values
are complex, not real, and the 
branch 0 real Lambert line 
(solid green curve) goes up and
to the right, starting from 
the point $(-1,0)$ (and shifted up by 0.3).

\textbf{Branch -1:}
We have used a dashed green
line for branch -1, and subtracted
$0.3$ from the value of $v$, so that we
are graphing
\begin{eqnarray}
  w &=& u + \mathbf{i}v 
                - \mathbf{i}[\text{offset}] \\
     &=& W_{-1}(z) - 0.3\mathbf{i}
\end{eqnarray}
for real $z = x$ for which $x < 0$.
This displays only $w$-plane image of
the negative real axial ray of the $z$-plane.
When $-1/e \le x < 0$, 
the branch -1 real Lambert line
for negative $x$
(dashed green line) is horizontal,
shifted 0.3 below the abscissa,
and that portion of the Lambert
line has $u < -1$.
When $x < -1/e$, the $w$ values
are complex, not real, and the 
branch -1 real Lambert line for
negative $x$ 
(dashed green curve) goes down and
to the right, starting from the point
$(-1,0)$ (and shifted down by 0.3).

Without the offsets, the dashed green
curves would be exactly the dashed red
curves, for the branch 0 real Lambert
lines, and for the branch -1 real Lambert
line for $x < 0$.

\subsubsection{How to Draw the Lambert Lines}

It is perhaps worthwhile describing how the
Lambert line curves can be produced using a
standard symbolic mathematics software package.
The curves in this document were drawn with
Maple, but other software with a Lambert $W$
function implementation will be similar.

The technique is to draw a Lambert $W$ line
as a parametric curve, with parameter $t$,
for instance.  Assume a branch number $k$.
Assume an axial ray direction in the $z$-plane,
denoted by $z_0$ equal to one of the four
directions from the origin: 
$+1, \mathbf{+i}, -1, -\mathbf{i}$.
Then let the parameter $t$ vary between
min and max limits, either linearly or
logarithmically as suits the graphical
requirements, and evaluate $w = W_k(t z_0)$.
This variable $w$ is a complex number.
Define $u,v$ as the real and imaginary parts
of $w$, and graph the parametric curve
$(u(t),v(t))$ for $t = t_{min} ... t_{max}$.

One further caveat: when $z \approx -1$,
the value of $w$ can change rapidly.
The real Lambert lines for branches
0 and -1 have a sharp corner at $z = -1$.
Hence special care is needed near
that point in the parametrization.
It is simplest to draw those Lambert lines
(branches 0 and -1, real Lambert lines
near $z = -1$)
as two distinct parametric curves,
one for $z \le -1$ and the other for
$-1 \le z < 0$, as each of those
curve segments is a smooth curve.

\subsubsection{Symmetry in the $w$-Plane}

Because the values of $v \cot(v)$ and $v \tan(v)$ 
are unchanged if $v$ is replaced by $-v$, the
Lambert lines (figure \ref{fig01}) are
symmetric above and below the horizontal axis.
Note however that the Lambert lines are not
symmetric about the vertical axis.

It is sufficient, in many geometric explorations,
to consider only the upper half of the $w$-plane.
The exception is when there are ancillary equations
in the problem which are not themselves symmetric
in $v$.  However, for an understanding of a system,
it can nonetheless be convenient to show all 
four quadrants of the relevant curves in the $w$-plane 
and the corresponding curves in the $z$-plane.
That leads to the two-planes method, which we will
discuss in the next subsection.

\subsection{What is the Two-Planes Method?}

The {\textbf{two-planes method}} involves considering
and comparing the solution curves for a problem, using
both the $w$-plane and the $z$-plane.  Recall that the
mapping from the $w$-plane to the $z$-plane is
$z = F(w) = w \, e^w$, and the mapping from the
$z$-plane to the $w$-plane is the multi-branch
Lambert $W$ function.

\subsubsection{Illustration: FSW with Strength $R=5$}

We illustrate with a 1-dimensional quantum FSW
of strength $R$.  Here $R$ is a dimensionless parameter,
determined by the depth and width of the FSW.
We will discuss the details of the factors which form
the strength $R$, in the next section \ref{sect3d}
of this paper.

\begin{figure}[ht]
	\centering
	\includegraphics[width=1.0\textwidth]{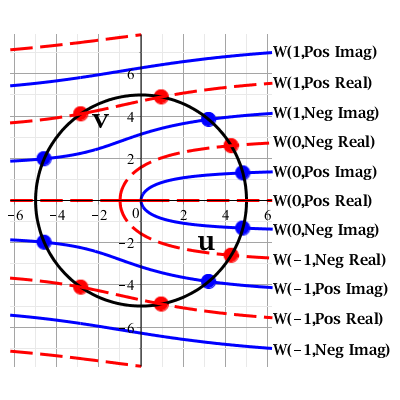}
     \caption{Lambert Lines and Strength R=5 Circle.
     This figure shows the Lambert lines in the
      $w$-plane, with an FSW circle of 
      strength $R=5.0$ superimposed.
      Dots have been placed to indicate where the
      strength circle intersects a Lambert line.
      The intersections of imaginary Lambert lines
      (blue solid curves) with the 
      strength circle are solutions
       of $u = v \tan(v)$ and $u^2+v^2 = 5^2$,
       and indicate even bound states of the FSW.
       The intersections of real Lambert lines
       (red dashed curves) with the 
       strength circle are solutions
       of $u = -v \cot(v)$ and $u^2+v^2 = 5^2$,
       and indicate odd bound states of the FSW.\\
     ------------------------------------------------------------------------------------------} 
	\label{fig06}
\end{figure}

Suppose that $R = 5$.
We wish to represent the simultaneous solutions
of equations (\ref{equvr}) and (\ref{equvtan}) 
in order to identify even bound states
of the FSW.
Draw a circle of radius $R=5$ about
the origin, in the $w$-plane.  
This is shown in figure \ref{fig06}.  
The blue imaginary Lambert lines
are the solutions of $u = v \tan(v)$,
so the intersections of those Lambert lines
with the $R=5$ circle identify the 
even bound state solutions of the FSW problem.
We see, from figure \ref{fig06},
that there are two even bound states of the FSW
in the first quadrant (where $u>0$ and $v > 0$)
and one bound state ``solution'' in the
second quadrant (where $u<0$ and $v>0$).

Similarly, the odd bound states of the FSW are
represented by the simultaneous solutions
of equations (\ref{equvr}) and (\ref{equvcot}). 
Those solutions are intersections of the $R=5$
circle with the red real Lambert lines, since those
lines are the solutions of $u = -v \cot(v)$.
We see, from figure \ref{fig06},
that there are two odd bound states of the FSW
in the first quadrant (where $u>0$ and $v > 0$)
and one bound state ``solution'' in the
second quadrant (where $u<0$ and $v>0$).

All those intersections, for even bound states
and for odd bound states, are marked in
figure \ref{fig06} with dots.
However, we omitted the $u = R, v = 0$
intersection since that is not a physically
meaningful solution for an FSW.

\subsubsection{Symmetry, Asymmetry, and $u < 0$}

Notice there is a symmetry among the
solutions.  Solutions for $v > 0$ are
paired with solutions for $v < 0$.
However, there is also an asymmetry:
Equation (\ref{equvr}) combined with
one of either 
equation (\ref{equvtan}) (for even states)
or equation (\ref{equvcot}) (for odd states)
has solutions for $u < 0$ which are not
in the first quadrant of figure \ref{fig06}.

Are solutions for $u < 0$ physically meaningful
for an FSW model?  
They are certainly mathematically meaningful,
as solutions of a pair of simultaneous equations
which appear in a mathematical model of a
physical system.
The answer to that question, whether solutions
with $u < 0$ are physically meaningful,
depends upon the specific structure 
of the physical system which has been 
described by a mathematical FSW-like model.
For the present, we will continue to include
the solutions with $u < 0$ in our explorations.

\subsubsection{Solutions in the $z$-Plane}

Now, let's ask what the corresponding 
solutions diagram looks like in the $z$-plane.  
The Lambert lines map into the axes of the
$z$-plane, with the even bound states
(on blue imaginary Lambert lines) going to
the positive and negative imaginary axes,
and with the odd bound states
(on red real Lambert lines) going to the
positive and negative real axes.

\begin{figure}[ht]
	\centering
	\begin{subfigure}[t]{0.45\textwidth}
	    \centering
	    \includegraphics[width=\linewidth]{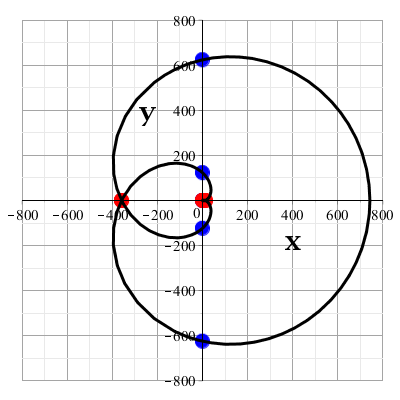}
          \caption{Complete Curve}
     \end{subfigure}
     \hfill
 	\begin{subfigure}[t]{0.45\textwidth}
	    \centering
	    \includegraphics[width=\linewidth]{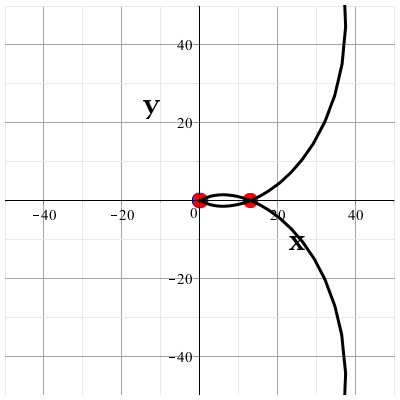}
          \caption{Magnified 16x}
     \end{subfigure}
     \begin{subfigure}[t]{0.45\textwidth}
	    \centering
	    \includegraphics[width=\linewidth]{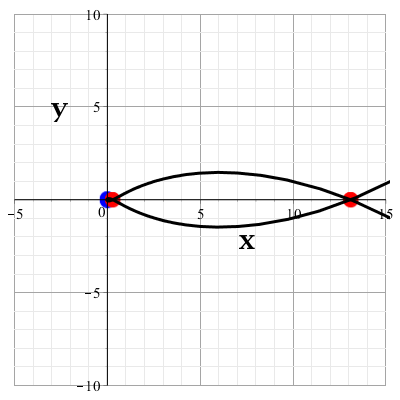}
          \caption{Magnified 160x}
     \end{subfigure}
     \hfill
 	\begin{subfigure}[t]{0.45\textwidth}
	    \centering
	    \includegraphics[width=\linewidth]{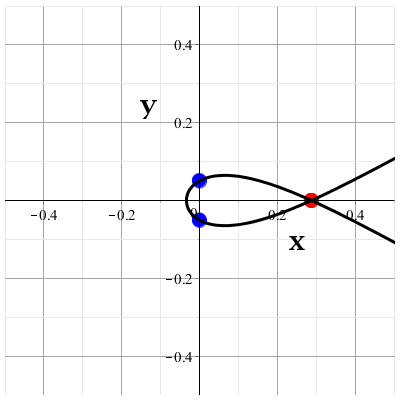}
          \caption{Magnified 1600x}
     \end{subfigure}
     \caption{$z$-Plane Image of Strength R=5 Circle.
     This figure shows the $z$-plane image of an
     FSW circle of strength $R=5.0$, at various
     magnifications in order to exhibit the detailed
     structure of the curve near the origin.
     The image curve is closed, just as the 
     strength circle in the $w$-plane is a closed curve.
     The Lambert lines of the $w$-plane correspond
     to the axial rays of the $z$-plane.
     Hence, wherever the strength circle image
     crosses the imaginary axis ($x = 0$) in the
     $z$-plane, there is an FSW even bound state
     indicated by a blue dot, and wherever the
     image crosses the real axis ($y = 0$) there
     is an FSW odd bound state, indicated by
     a red dot. \\
      ------------------------------------------------------------------------------------------} 
	\label{fig07z}
\end{figure}

The circle $u^2 + v^2 = R^2$, which is
a continuous closed curve in the $w$-plane,
will go to a continuous closed curve in the
$z$-plane, with multiple loops around the origin.  
See figure \ref{fig07z}.
There are large scale changes in the $z$-plane
curve, because of the $\exp(u)$ factor seen
in equation (\ref{eqzuv}).
Therefore the $z$-plane image of the strength
circle for $R=5$ has been drawn in full in
figure \ref{fig07z}a, and at various zoomed-in
magnifications in figures \ref{fig07z}b,
\ref{fig07z}c and \ref{fig07z}d.

It is easy to draw that curve in the $z$-plane.
Write $w = R \exp(\mathbf{i}\theta)$,
where the angle $\theta$ goes from 
$0$ to $2 \pi$, 
and walk around the $|w| = R$ circle 
in the $w$-plane, 
plotting $z = w e^w$ in the $z$-plane.
As can be seen in the various magnified
views in figure \ref{fig07z},
the blue dots, which indicate odd bound
states of the FSW, all lie upon the imaginary
axis ($x=0$), and the red dots, which
indicate even bound states of the FSW,
all lie upon the real axis ($y=0$).
No dot was drawn at the point $(R,0)$
in the $z$-plane, because that is not a
physically feasible state of the FSW.

The two-planes method provides two ways
of viewing the properties of an FSW-like
system.  
In the $w$-plane, the Lambert lines, which
represent the algebraic relationships given
by equations (\ref{equvtan}) and 
(\ref{equvcot}), are
complicated curves, requiring one to be
able to calculate Lambert $W$ function
values, or at least to be able to calculate
values of the tangent function.
But in the $z$-plane, the Lambert lines
are simply the imaginary and real axial
rays, and can be drawn with a straightedge.
On the other hand, the equation (\ref{equvr})
which represents the strength of an FSW,
is just a circle about the origin in the
$w$-plane, and can be drawn with a 
compass.  But in the $z$-plane, one has
to draw the image of that strength circle
by calculating an exponential, and also 
has to deal with changes of scale in the
diagram.

Thus the two-planes method provides one
with a choice of representations to use for
solving a physical problem.  As is familiar
from other mathematical physics problems,
the choice of a convenient coordinate system
or representation of the physical system can
make a problem either fairly easy to solve
or  uncomfortably difficult.

To assist in keeping track of the properties
of the two types (imaginary and real) 
of Lambert lines, the colours and line
styles used for those curves, 
and the two types of FSW bound states 
(even and odd) which those lines correspond to, 
we have appended Table 1 which records 
all that information on a single page.



\section{3-D Representation of Sensitivity of Solutions}
\label{sect3d}

In this section, we wish to explore some
techniques which are relevant to determining
special solutions of a physical system.
For example, a tangency which may suggest
desirable parameters for a physical system
used as a sensor, or alternatively may suggest
an operating context in which a physical system
may be so sensitive to parameter changes that
it is prone to failure.
We also wish to describe a method
of introducing a third dimension into the
$w$-plane and $z$-plane representations
of the system, in order to visualize situations
in which the behaviour of the system can
change when a parameter is altered. 

We will use a finite square well as our illustrative
system.  In this illustration, we will ignore the
question of whether solutions with $u < 0$ are
physically meaningful.  We are allowing the FSW
to be a proxy for an FSW-like system, in which
we are solving for intersections between a
strength circle of radius $R$, and some Lambert
lines.

What happens if the strength of the FSW changes?
That may happen because the depth of the well
changes, or because the width of the well
changes.  Those terms, depth and width, have a
meaning appropriate to the particular physical
system which is being modelled via an FSW.
The depth, for instance, may be related to 
a bias voltage applied to a sensor.
Since the FSW discussed here as an illustration
is merely a proxy for other FSW-like systems, 
one can think of any other parameter, 
not necessarily strength of an FSW, 
which is an adjustment for the system, 
either at design or during operation.

\subsection{Strength Formula for a Finite Square Well}

Consider a 1-dimensional quantum finite square well.  In a standard representation, 
found in quantum mechanics books,
the FSW is described in terms of 
a width $L$ (a length)
and a potential depth $V_0$ (an energy).
For instance, an FSW might be 
stated to have a potential function
$V(x)$ which is a fixed negative energy level, 
$V(x) = -V_0$ for $x$ satisfying $-L < x < L$.
Outside the well, 
the potential function $V(x) = 0$.

We call $L$ the half-width of the well, 
and $V_0$ (positive) the depth of the well.  
The energy $E$ of a bound state will be 
negative, and will satisfy $-V_0 < E < 0$.
There are only a finite number of possible bound
state energies for the FSW which will produce a
stationary wave function which satisfies the  
{\Schrodinger} equation.

We will not go through the details of finding the
solutions here.\footnote{In references
\cite{RV2014,RV2017} we show the details
of using Lambert lines to find the bound states
of a one-dimensional quantum finite square
well.}
Here, we simply wish to exhibit the
formulas which determine the FSW strength
$R$, and to indicate how the $u$ and $v$
values are determined once the bound state
energy is known.

The relevant formulas are
\begin{eqnarray}
  R &=& \frac{L}{\hbar} \, \sqrt{2m V_0} \\
  u &=& \frac{L}{\hbar} \, \sqrt{2m |E|} \\
  v &=& \frac{L}{\hbar}  \, \sqrt{2m (V_0 - |E|)}
\end{eqnarray}
Here $m$ is the effective mass of the bound particle.
Recall that $V_0$ is positive, 
that the bottom of the potential well
is at $V(x) = -V_0$, and
that the bound state energy $E$ is 
a negative value between 0 and $-V_0$.
The symbol $\hbar$ is of course the 
reduced Planck-Einstein-Bose (PEB)
constant.\footnote{The PEB constant
is customarily called Planck's constant.
We prefer ``PEB constant''.}
The positive signs assigned to $u$ and $v$
by the above formulas are conventional,
and we are at liberty, depending upon 
the specifics of the system described via
an FSW, to take the negative
signs for the square roots.
The FSW analytical work which leads to
those formulas simply requires that 
$u, v$ satisfy
\begin{eqnarray}
  \label{eqfswusq}
  u^2 &=& \frac{2 m |E| L^2}{\hbar^2} \\
  \label{eqfswvsq}
  v^2 &=& \frac{2 m (V_0 - |E|) L^2}{\hbar^2}
\end{eqnarray}
and also requires that $u,v$ be related
by either $u = v\tan(v)$ or $u = -v\cot(v)$.
That $R$ satisfies equation 
$u^2 + v^2 = R^2$ follows by adding
together the two equations
(\ref{eqfswusq}) and (\ref{eqfswvsq}).
All three of the quantities $u,v,R$ are
dimensionless.

If $v = 0$, then $E = -V_0$, which means,
in a quantum FSW, that the bound state
energy would lie exactly at the bottom of
the well.  That is unphysical,
since such an energy would violate the
uncertainty principle of quantum theory.
For that reason, $v=0$ is excluded from
the set of allowable solutions which we
discuss here.  However, there may be
systems, classical perhaps, for which the mathematical solution $v=0$ might make 
physical sense.
Here, for simplicity, and because the
solutions with $v=0$ are not as interesting,
we will assume that $v \ne 0$.

If $u=0$, then $E=0$, which means that
the state with energy $E$ is not actually
a bound state.  The FSW potential outside
the well is assumed to be zero, and a
particle with energy $E$ would be
uncertainly free, or uncertainly bound.
That is a different situation
than the system states
which we are contemplating here,
but suggests some physically interesting
opportunities.  For instance, scattering or
resonance/delay of a free particle with 
low positive energy which passes over
a potential well.

\subsection{Tangency of Lambert Line and Strength Circle}

We wish to explore the possibility
that neither of $u,v$ is zero, and that
the strength of the FSW is one of the
special values which makes the strength
circle tangent to a Lambert line.
It is the hypothesis of this paper that such
tangencies may be useful for designing sensors
which utilize physical systems
which are FSW-like.  That is, these systems
may be ones which have equations similar
to equations (\ref{equvr}) to (\ref{equvcot})
to be solved, and for which $u < 0$ cases
are meaningful in the model of the system.
Our objective in this paper is to describe
ways of working with such systems, using
Lambert lines techniques.

\begin{figure}[ht]
	\centering
	\includegraphics[width=1.0\textwidth]{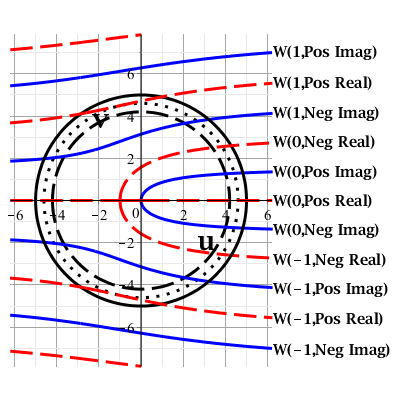}
     \caption{Lambert Lines and Three FSW Strengths.
     This figure shows the Lambert lines in the
      $w$-plane, with FSW strength circles of 
      three different strengths superimposed.
      The solid circle is strength $R=5.0$,
      which was previously shown in 
      figure \ref{fig06}.
      It intersects a Lambert line six times in
      the upper half-plane ($v>0$).
      The dashed circle is strength $R=4.2$
      and it intersects a Lambert line only
      four times in the upper half-plane, 
      so there are fewer bound states
      for the FSW.
      The dotted circle is strength $R=4.6034$
      and it intersects a Lambert line 4 to 6 times.
      (The number of intersections is uncertain.)
      Two of those intersections, at $u=-1$
      and $v \approx 4.4934$ coincide,
      where the strength circle is tangent
      to a Lambert line.
      An FSW with strength 4.6034 may be
      a candidate for design of a sensor.  \\
     ------------------------------------------------------------------------------------------} 
	\label{fig08}
\end{figure}

With $R=5$, we observed in figure \ref{fig06}
that there are six bound states in the upper two
quadrants ($V > 0$) of the $w$-plane.
Let's draw some more strength circles.
See figure \ref{fig08},
where the strength $R=5$ circle is represented
by a solid curve.
With strength $R=4.2$, represented by a dashed
circle in the same figure,
there are only four bound states in the upper
half-plane.
At strength $R$ slightly more than $R=4.6$,
for example at $R=4.6034$ 
represented by a dotted circle, 
the circle has become tangent
to the blue Lambert line and
there are an uncertain number of bound
states in the upper half-plane:
maybe four bound states,
or maybe six bound states,
or maybe (on average) five bound states.

The strength $R = 4.6034$ for the FSW
thus represents a possible sensitivity, 
which is a guideline for the design
of a sensor or an inherently unstable
system. 
Two of the odd bound states are
at almost the same phase angle.
However, the phase angle can 
change by a few percent while the 
radius of the strength circle changes
by only a small fraction of a percent.
That is the sensitivity in the system. 
A slight change in the physical properties
of the system being modelled via an FSW, 
perhaps due to flexure, or incoming energy, 
or a change in a bias voltage, 
may produce a change in the behaviour 
of the system.    For instance, there may
be an increase in the system's specific
heat, or a release of energy which has
been pumped into a system which is at
a quasi-stable equilibrium.

One can see that the energy level associated
with a point of tangency is not the same as
the energy level at which the FSW strength
circle intersects the positive $v$-axis.
Let $(u_1,v_1)$ denote the point of tangency
when $R \approx 4.6034$.
Then $u = -1$, and $v$ can be calculated
from equation (\ref{equvr}) to be
$v \approx 4.4935$.

If one were designing a sensor with attention
focused only on the system's behaviour in
the first quadrant of the $w$-plane, where
$u>0$ and $v>0$, then one might decide 
to set the system's strength to $R = 3\pi/2$
or approximately 4.7124.
That is about 2.3 percent more than the
value of $R$ which corresponds
to greatest sensitivity, if the system is one
for which values $u<0$ have a physical
meaning.

\subsection{How to Find Points of Tangency}

We wish to describe how to find $w$-plane
points where the strength circle of an FSW
is tangent to a Lambert line.

\textbf{Fact 1: Tangency requires $u < 0$.}

The first observation is that tangencies
arise only in quadrant 2 or 3 of the $w$-plane,
when $u < 0$.
That may be seen geometrically.  We can 
restrict our verification to the upper half-plane,
where $v>0$, because the lower half-plane
is a mirror image of the upper half-plane.
Moreover, we can exclude the particular case
of tangency to the real Lambert lines which
pass through $(-1,0)$, because $v=0$
there (and because there is a sharp corner in the
Lambert line curves at that point).

Each of the other Lambert lines in the 
upper half-plane is an increasing function.  
To see this, note that $u = v\tan(v)$ increases
as $v$ increases within a horizontal band of
values of $v$ which lies between two positive odd
integer multiples of $\pi/2$ (see figure \ref{fig03})
so those Lambert lines go from the lower left
asymptote of the band to the upper right
asymptote of the same band. 
And similarly, $u = - v\cot(v)$ also increases
within its band of asymptotes, from one even
positive integer multiple of $\pi/2$ to the next,
from lower left in the band to upper right.

Imagine a particular point $(u_1,v_1)$
on an upper half-plane Lambert line.
A perpendicular from that point must 
descend down to the right, and hence 
intersect the $v=0$ axis at a point
$(u_2,0)$ for which $u_2 > u_1$.
For the Lambert line to be tangent
to a circle about the origin, that
perpendicular must pass though
the origin, so we require $u_2=0$
for tangency.  
Hence $u_1 < 0$, which means that
the Lambert line's point of tangency
must lie in quadrant 2.

\textbf{Fact 2: If $(-1,v)$ lies on a Lambert
line, then a strength circle of radius
$R = \sqrt{1+v^2}$ will be tangent
to that Lambert line.}

Suppose that ($u_1,v_1)$
lies on a Lambert line, and $u_1 = -1$
and $v_1>0$.
We wish to show that the straight line
from $(-1,v_1)$ to the origin will be
perpendicular to that Lambert line.
The slope of a Lambert line is $dv/du$,
and the slope of a radius of a circle
from the origin is $v/u$.  To have 
the Lambert line be perpendicular
to the radius, the product of those
two slopes must be -1.  
That is, we wish to verify that
\begin{eqnarray}
    \frac{dv}{du} \cdot \frac{v}{u} &=& -1 ,
\end{eqnarray}
or equivalently,
\begin{eqnarray}
    \frac{du}{dv} \cdot \frac{u}{v} &=& -1 ,
\end{eqnarray}
at the point $(-1,v_1)$.

As before, we proceed by cases.

Case 1: Suppose that the Lambert line
in question is an imaginary Lambert line,
so that $u = v\tan(v)$ is the
equation of the Lambert line.
Multiply each side of $u = v\tan(v)$
by $\cos(v)$ to obtain $u\cos(v) = v\sin(v)$.
Differentiate both sides, to obtain
\begin{eqnarray}
    du \, \cos(v) - dv \, u \sin(v) &=& 
    dv \, \sin(v) + dv \, v \cos(v)
\end{eqnarray}    
Transposing and simplifying, we get
\begin{eqnarray}
    du \, \cos(v)  &=& 
    dv \, [(u+1) \sin(v) + v \cos(v)] \\
    \frac{du}{dv} &=&
    \frac{(u+1) \sin(v) + v \cos(v)}{\cos(v)} \\
    \frac{du}{dv} &=&
    (u+1) \tan(v) + v
\end{eqnarray}
Since $u = v\tan(v)$, this last equation
becomes
\begin{eqnarray}
\label{eqdudv1}
  \frac{du}{dv} &=& \frac{u(u+1)}{v} + v
\end{eqnarray}
Multiplying by $u/v$ gives
\begin{eqnarray}
\label{eqdudv2}
  \frac{du}{dv} \cdot \frac{u}{v} &=& 
     \frac{u^2 (u + 1)}{v^2} + u
\end{eqnarray}

Now suppose that $u = -1$.  Then the first
term in the last expression disappears,
and the second term is $-1$, which 
means that the radius of the strength
circle is tangent to the Lambert line.
That completes the proof of case 1.

Case 2:  Suppose that the Lambert line
in question is a real Lambert line,
so that $u = -v\cot(v)$ is the
equation of the Lambert line.
Quite similar to case 1.
One ends up with equations (\ref{eqdudv1})
and (\ref{eqdudv2}) as in case 1,
and then taking $u=-1$ completes
the proof.

\begin{figure}[ht]
	\centering
	\begin{subfigure}[t]{0.45\textwidth}
	    \centering
	    \includegraphics[width=\linewidth]{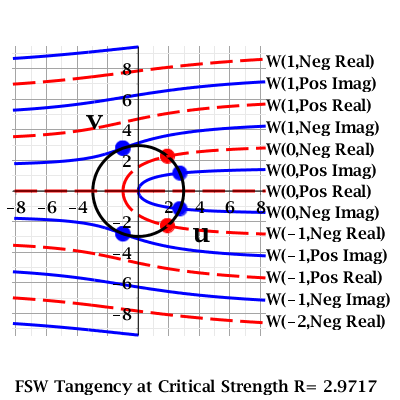}
          \caption{Tangency at $(-1,2.7984)$}
     \end{subfigure}
     \hfill
 	\begin{subfigure}[t]{0.45\textwidth}
	    \centering
	    \includegraphics[width=\linewidth]{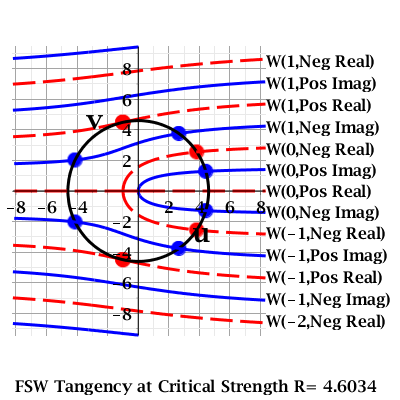}
          \caption{Tangency at $(-1,4.4934)$}
     \end{subfigure}
     \begin{subfigure}[t]{0.45\textwidth}
	    \centering
	    \includegraphics[width=\linewidth]{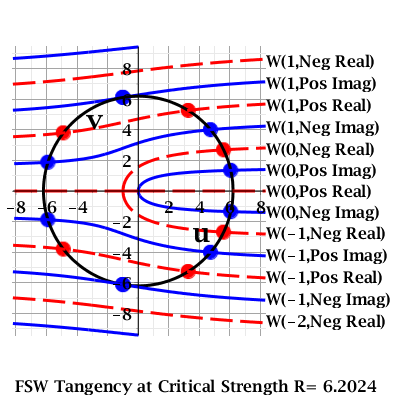}
          \caption{Tangency at $(-1,6.1213)$}
     \end{subfigure}
     \hfill
 	\begin{subfigure}[t]{0.45\textwidth}
	    \centering
	    \includegraphics[width=\linewidth]{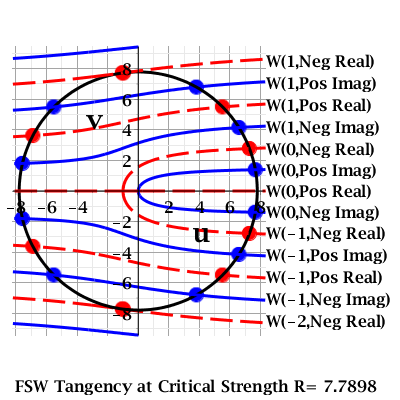}
          \caption{Tangency at $(-1,7.7253)$}
     \end{subfigure}
     \caption{Tangencies of the FSW Strength
     Circle.
     This figure shows the tangency of an FSW
     strength circle to a Lambert line, for four
     different values of the strength $R$.
     Each point of tangency lies upon the
     vertical line $u=-1$. \\
      ------------------------------------------------------------------------------------------} 
	\label{fig09z}
\end{figure}

Figure \ref{fig09z} shows the FSW bound
state solutions for the first four critical
strengths: 
$R = 2.9717, 4.6034, 6.2024, 7.7898$.

\begin{figure}[ht]
	\centering
	\includegraphics[width=1.0\textwidth]{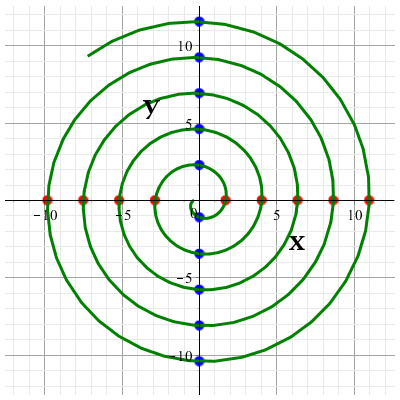}
     \caption{$z$-Plane Image of $u=-1,v>0$ Line.
     This figure shows the image, in the $z$-plane,
     of the vertical line $u=-1, v>0$ in the
     $w$-plane.  Dots have been placed where
     the curve crosses one of the axial rays.
     These correspond to intersections of the
     vertical line $u=-1$ with Lambert lines
     in the upper half of the $w$-plane, 
     which are points where an 
     FSW strength circle of a critical radius
     will be tangent to a Lambert line.   \\
     ------------------------------------------------------------------------------------------} 
	\label{fig10}
\end{figure}

In order to find the critical strengths $R$,
we are essentially moving up along
the vertical line $u=-1$ in the $w$-plane.
A very interesting diagram results
if we map the line $u=-1,v>0$ to the
$z$-plane.  It is a spiral (see figure \ref{fig10}), which intersects one of the four axial rays in the
$z$-plane at $z$-values which can be
used to determine the $R$-values which
are the critical strengths.
Given a $z$-value, we can calculate
$w=W_k(z)$ for the appropriate branch
number.  The $v$ coordinate of the
point of tangency in the $w$-plane is
the imaginary component of $w$.
The critical strength $R$ can then be
calculated from $v$.
Moving back to the  $z$-plane, 
we can determine the magnitude 
of the corresponding $z$-value,
via
\begin{eqnarray}
  |z| &=& |w|\exp(u) = \frac{|w|}{e} 
       = \frac{R}{e}.
\end{eqnarray}

That, of course, gives a formula which
can be used directly, $R = e\,|z|$,
for determining the critical radius from
a knowledge of the $z$ values for
which the $z$-plane image of the
$u=-1$ line from the $w$ plane
intersects one of the axial rays
in the $z$-plane.
There is no need to determine the branch
number of the Lambert $W$ function in
order to find the critical $R$ values.
The consecutive critical $R$-values 
on the vertical line $u=-1,v>0$
in the $w$-plane, 
are spaced about $\pi/2$
apart.
The magnitudes of their $z$-plane
images are spaced about 
$\pi/2e \approx 0.5779$ 
apart in the $z$-plane.
Those $z$-values are distributed
among the four axial rays from the
origin, so along each axial ray the 
magnitudes of the $z$ values are
spaced about $2\pi/e \approx 2.3115$
apart.

Figure \ref{fig10} shows the $z$-plane
image of the $w$-line $(-1,v)$ for $v > 0$.
Dots have been placed on the axial rays
for the first 19 critical strengths $R$.

\subsection{3-Dimensional Representation of Tangency}

The original impetus for preparing this document
was a desire to extend the two-planes method to
a 3-dimensional representation, in order to use
geometric methods to visualize systems which
are FSW-like, that is, 
which include equations such as
(\ref{equvtan}) and (\ref{equvcot}), and which
also include a parameter such as the strength $R$
of an FSW, or the refractive index $n$ of a
dielectric in a waveguide.

Conceptually, one can think of the four FSW
solutions shown in figure \ref{fig09z} as slices 
of a 3-dimensional space, with different values
of the FSW strength $R$.  
Thus $R$ is the third dimension in a visualization.
The curved Lambert lines become curved sheets,
call the Lambert sheets for convenience.
The surface which represents the strength
circles $u^2 + y^2 = R^2$ in three dimensions
is a cone, with its point at the origin
$(u,v,R) = (0,0,0)$.
Simultaneous solutions of equation (\ref{equvr})
and one of equations (\ref{equvtan}) or
(\ref{equvcot}), are the curves where the
FSW strengths cone intersects the Lambert 
sheets.

Our first attempts to demonstrate this geometry, 
by drawing the two types of surfaces and
highlighting their intersections, were not
satisfactory.  The diagrams have too much
clutter to allow clear reasoning about the
situation.

Instead, we have chosen to show only the
lines of intersection of the surfaces.
One must imagine the Lambert sheets and
the FSW strengths cone whose intersections
produce those intersection lines.

\begin{figure}[ht]
	\centering
	\begin{subfigure}[t]{0.45\textwidth}
	    \centering
	    \includegraphics[width=\linewidth]{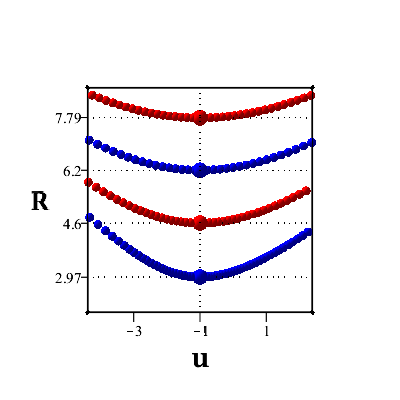}
          \caption{Projected along $v$ axis.
             Shows that critical points 
                occur at $u=-1$.}
     \end{subfigure}
     \hfill
 	\begin{subfigure}[t]{0.45\textwidth}
	    \centering
	    \includegraphics[width=\linewidth]{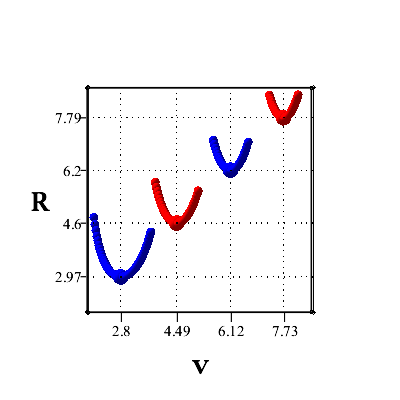}
          \caption{Projected along $u$ axis.
             Shows relationship of $v$ and $R$
               at critical points.}
     \end{subfigure}
     \begin{subfigure}[t]{0.45\textwidth}
	    \centering
	    \includegraphics[width=\linewidth]{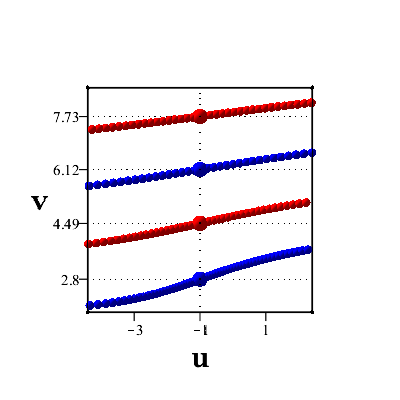}
          \caption{Projected along $R$ axis.
             Shows relationship of $u$ and $v$
               at critical points.}
     \end{subfigure}
     \hfill
 	\begin{subfigure}[t]{0.45\textwidth}
	    \centering
	    \includegraphics[width=\linewidth]{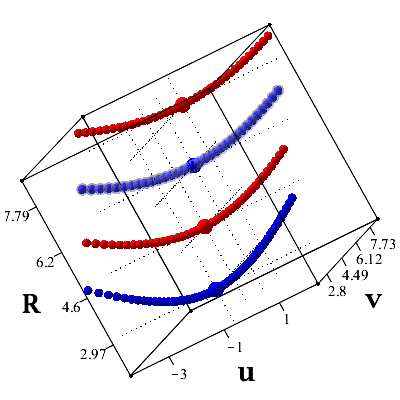}
          \caption{Tipped to show all 3 axes.
             Demonstrates Lambert lines
               as 3-D curves.}
     \end{subfigure}
     \caption{Three-Dimensional View of
        Intersections of Lambert Line Sheets
        and the Cone of FSW Strength Circles.
        All points represent a bound state of
        a finite square well of strength $R$.
        Critical points are tangencies which
        possibly represent good sensor
        configurations. \\
      ------------------------------------------------------------------------------------------} 
	\label{fig11z}
\end{figure}

Figure \ref{fig11z} has some views of the
FSW lines of intersection.
Each figure \ref{fig11z}a to \ref{fig11z}d
shows a different view of the 3-D lines
of intersection.

In figure \ref{fig11z}a, we see a projection
along the $v$ axis, so that this figure is
essentially a graph of $R$ vs $u$.
There are four curves of intersection shown,
because four Lambert sheets have been
included in the diagram.
The $R$ value along the vertical axis of the
diagram has a maximum of 8.7, 
and the $v$ values are given by 
$v = \sqrt{R^2 - 1}$, which thus has a
maximum value of about 8.65, less than
$3\pi$, so there are only four Lambert
sheets of interest.  (We have omitted the
Lambert sheets which do not produce a
tangency with the $R$-cone.)
The number of intersection lines would be
more if the diagram had been drawn to
include a larger maximum value for $R$.
Faint dotted lines have been included to
pass through the $R$-minima of each
of the curves shown, with tick marks
along the vertical $R$-axis to illustrate
that those minima correspond to the
critical $R$-values exhibited in figure \ref{fig09z}.
One can also see, in figure \ref{fig11z}a,
that all the critical $R$-values correspond
to $u=-1$.

In figure \ref{fig11z}b, we see a projection
along the $u$ axis.
There are again four curves of intersection shown.
Faint dotted lines pass through the critical points.
The critical FSW strengths $R$ are 
marked on the vertical axis, and
the corresponding values of $v$ are
marked on the horizontal axis.

In figure \ref{fig11z}c, we see a projection
along the $R$ axis.
This shows the relationship of $u$ and $v$,
which of course is the same curves that
are shown in a 2-dim Lambert lines diagram.
The difference is that the 3-dim curve, if
rotated to show the $R$ axis, would display
$(u,v,R)$ points where $R = \sqrt{u^2+v^2}$.

In figure \ref{fig11z}d, the 3-dim diagram
has been rotated to show all three axes.
Working with this, and similar, diagrams
in a 3-dim graphics software package can
be an aid to understanding the bound states
of an FSW.

It is our belief that 3-dim methods such
as have been used in this section for a
finite square well, can also provide insight
into the properties of FSW-like systems.
In future reports, 
we wish to extend the techniques used
above, to other FSW-like systems.



\section{Discussion and Conclusions}
\label{sectdisc}

In this paper, we have shown that the
geometric display of the bound state
energies of a 1-dimensional finite
quantum square well (FSW) provides
a convenient visualization of FSW strengths
which can be useful for designing sensors.
As well, we have exhibited a 3-dimensional
visualization of the bound state solutions,
and of the critical strengths of an FSW.





\begin{table}[ht]
\centering
\begin{tabular}{|| p{6cm} | l | l ||}
\hline \hline
 & & \\
\textbf{Type of}
  &  & \\  
\textbf{Lambert Line} 
  & \textbf{Imaginary} 
  & \textbf{Real} \\
 & & \\
\hline \hline
 & & \\
Equation in $w$-plane
  & & \\
where $w = u + \mathrm{i}v$ 
  & $u = v \, \tan(v)$
  & $u = -v \, \cot(v)$ \\ [1mm]
\hline
 & & \\
Intersects ordinate
  & & \\
at $v = \pi/2$ times an
  & EVEN integer 
  & ODD integer \\ [1mm]  
\hline  
 & & \\
Asymptotes cross ordinate
  & & \\
at $v = \pi/2$ times an
  & ODD integer 
  & EVEN integer \\   [1mm]
\hline  
 & & \\
Shown in figures as
  & BLUE SOLID
  & RED DASHED \\ [1mm]
\hline  
 & & \\
Axial ray in $z$-plane
  & $x$ is equal to 0
  & $x > 0$ or $x < 0$ \\
where $z = x + \mathrm{i}y$ 
  & $y < 0$ or $y > 0$ 
  & $y$ is equal to 0 \\ [1mm]
\hline  
 & & \\
FSW bound state 
  & & \\
wavefunctions are
  & EVEN functions 
  & ODD functions \\  [1mm] 
\hline \hline
\end{tabular}
\label{tab01}
\caption{Lambert Line Properties and Notation}
\end{table}

\end{document}